# Who May Release, Who Must Repair: Separating Authorization from Report Accountability in Personalized Services

*Short Paper*

**Chang Xu**
Independent Researcher
Seattle, US
changxu109@gmail.com

**Junyuan Tan**
Independent Researcher
Burlingame, US
junyuan_tan@berkeley.edu

## Abstract

*Personalized services often rely on different parties to compute results and decide whether data permissions and service policies allow delivery. The computing party's report may omit permitted inputs, leaving the delivery party to decide whether to wait for correction. Requiring correction first can delay service despite valid input permissions. Delivering first instead requires preserving responsibility for the incomplete report. We design an interface that checks input permissions separately from reporting requirements and records who must correct omissions and who may verify corrections. We evaluate three implementations on scripted histories and generated cases against a design treating omissions as authorization failures. When policy permits delivery before correction, our interface releases results that the alternative withholds while preserving repair assignments. These assignments remain in force through policy changes, so changing delivery rules does not erase earlier obligations. The interface makes the basis for release and outstanding repair responsibilities traceable across cooperating parties.*

## Introduction

Personalized services often distribute computation and delivery across teams or organizations. A music service, for example, may have one component rank songs and another check whether data permissions and service policy allow the ranking to be delivered. The ranking component submits an account of its inputs to support that decision. The delivery component also retains receipts identifying which resource responses were supplied to the execution.

Suppose the ranking component receives a listener profile and catalog data but lists only the profile in its report. If both inputs were permitted for personalization, the report needs correction even though the received inputs were authorized. The ranking may already be ready for delivery. Requiring correction first then makes the listener wait for the missing catalog entry to be added to the report, although no new ranking needs to be computed. Yet delivering every available ranking without checking its inputs could release a result incorporating data the user did not permit for personalization. The service therefore needs to distinguish waiting for input permission to be established from waiting for an omitted report entry to be added. If delivery proceeds before correction, someone must remain responsible for correcting the report, and someone must decide whether the correction resolves the defect.

We call the component that computes and reports the **executor**, and the party that decides on delivery the **publisher**. To check permission independently of that report, the publisher compares its retained receipts with permission state supplied by a **consent authority**. An omitted input therefore remains part of the permission check. A **service owner** sets the reporting policy that determines whether an incomplete account delays delivery. Throughout this paper, report completeness means listing every resource and version in the received-input receipts.

The distinction between permission and report completeness must also survive subsequent decisions. The service owner may change from allowing delivery with repair pending to requiring correction first. A new release then needs the applicable policy and renewed permission evidence. An earlier defect, however, remains assigned to its original executor and to the party authorized to accept its correction, which we call the **closure authority**. The design challenge is to keep a current release decision from silently closing or transferring an earlier reporting obligation.

The setting connects to data governance as an allocation of decision rights (Khatri & Brown, 2010). It also arises in services that separate useful functions from the data flows supporting them. Miller et al. (2026) make this distinction in their study of online tracking. In personalization, Mosaic supplies reusable user representations to multiple consumers through online and offline paths (Zheng et al., 2026), while measurements of web consent systems document gaps between revocation interfaces and downstream communication (Kancherla et al., 2025). These studies motivate checking evidence across service boundaries. Our focus is how parties coordinate delivery and correction when that evidence differs from the executor's account.

We ask: **How can a service distinguish input authorization from report completeness when making release decisions, while preserving repair responsibility across later releases and policy changes?** We design an interface linking each release to its permission and policy evidence, and each defect to its original repair assignment. Three implementations and a controlled comparison evaluate this allocation through execution histories and generated cases. The contribution is an executable account of which decision permits delivery and which obligation remains open afterward.

## Foundations and Design Requirements

Design-science research develops and evaluates artifacts for relevant organizational problems (Hevner et al., 2004). Here the problem is allocating release and repair decisions when different parties supply permission evidence, received-input records and execution reports. Usage control connects authorization to mutable attributes and continuing use (Park & Sandhu, 2004). Consent monitoring checks collection or access events against evolving permission (Robol et al., 2023). These approaches establish why a decision needs an identifiable permission state. Our reporting-policy question follows once the received inputs are authorized: how should an incomplete account affect delivery and subsequent correction?

TrustDS and C-Trace connect policy to runtime decisions in complementary ways. TrustDS guards execution and release boundaries, recording the consent epoch, policy version and other runtime evidence supporting an allow or deny decision (Dockara & Malhotra, 2026). C-Trace evaluates predicates over agent events and records whether each event is forwarded, redacted or blocked (Kahani et al., 2026). We examine an additional relationship between a delivery decision and an obligation that can outlast it. A release may be permitted while its report still requires correction, and changing the reporting policy must preserve the authority assigned to that earlier defect.

This relationship gives the allocation an organizational meaning beyond an event log. The executor must provide a corrected account, whereas the closure authority decides whether the obligation has been fulfilled. Schmidt et al. (2025) show that evidence supporting accountability arguments affects AI developers' perceived accountability. Our records specify the evidence and roles that such an account can refer to. Whether practitioners understand and use this allocation is a separate question for the planned study.

Table 1 summarizes the requirements and the observations used to evaluate them. Case identifiers are retained in the reproducibility package. The main text uses them only where they help connect a concrete history to its evidence.

| Requirement | Design principle | Evaluation evidence |
|---|---|---|
| A report may omit a received input. | Check permission against retained input evidence. | Route, purpose and version contrasts. |

| Requirement | Design principle | Evaluation evidence |
|---|---|---|
| An authorized omission can support two delivery policies. | Separate permission from reporting requirements. | Matched release-with-repair and complete-report histories. |
| Correction can outlast delivery. | Preserve the original correction and verification assignments. | Repair after a policy change and an incomplete repair attempt. |
| The service makes another release decision. | Require renewed evidence for that decision. | Fresh versus stale decision bases and an event-order probe. |

**Table 1. Requirements, design principles and evaluation evidence.**

## Decision and Repair Interface

### *Input Evidence and the Delivery Choice*

The reference service makes the music example inspectable by ranking three items, A, B and C. Catalog resource c0 returns scores (3, 2, 1), and profile resource p0 returns (0, 0, 5). Adding them gives (3, 2, 6), ranked C, A, B. An advertising resource x0 returns (0, 4, 0), which would change the ranking if included. These small vectors make the input roles easy to follow. The final evaluation stage substitutes scores from a trained recommender.

A **resource** is a named input that returns a score vector. For each execution, the publisher retains a receipt identifying the resource and version received. These receipts form the **inventory** used to check report completeness. A separate permission record identifies the user, request purpose, resource, resource version and current version of permission state, called the **consent epoch**. An **authorization binding** links these fields so that the publisher can check whether the received version was permitted for this request.

Version agreement matters for cached inputs. A receipt identifying profile version 2 cannot establish permission from a record covering only version 1. The interface leaves authorization unresolved until the received version and its authorization evidence agree. This condition is distinct from a reporting omission: correcting the account does not itself supply the missing version-specific permission evidence.

In the example, the user has personalization and advertising permissions at epoch 1. The request's **route**, meaning its allowed set of input resources, includes p0 and c0. Profile p0 depends on feature f0, so checking p0 also requires checking f0. The interface checks these dependencies, route membership and permission for the request's purpose. Advertising permission alone therefore does not admit x0 to this personalization request.

When p0 and c0 are received but the report names only p0, the inventory is authorized and the report is incomplete. Under **release-with-repair**, the publisher delivers the ranking while the defect remains pending. Under **complete-report**, it holds the ranking until correction. Receiving x0 instead produces a hold under either policy. The policy choice thus governs what an authorized omission means for delivery.

### *Release Evidence and Repair Responsibility over Time*

The interface links the evidence in a sequence of records. An **input bundle** groups the inventory and original report with references to the inputs. A **permission snapshot** records the state checked for authorization. A **policy application** then identifies that authorization and the reporting policy in force. Comparing report and inventory produces a discrepancy record linked to the policy application. Finally, a **publication decision** cites these records and states whether to release or hold the output. Corrections and verification outcomes are appended to the same history, preserving the original report.

The service owner specifies the policy's scope, effective period and allocation of release, repair and closure rights. The publisher records the delivery decision. The executor proposes a correction for the assigned closure authority to verify. If a correction remains incomplete, its unsuccessful verification is recorded and the defect stays pending. A later policy does not reassign that defect's executor or closure authority.

Each later release also needs evidence prepared for that decision. Its policy application refers to the previous publication and a newly accepted permission snapshot, input bundle and authorization result. The permission contents may be unchanged, but the records establish that the check was made again. One check rejects a decision basis dated before the previous publication. A second requires the supporting records to have entered the history after that publication, so changing timestamps alone cannot refresh old evidence. An exact retry adds no record. These rules preserve a current authorization basis alongside any outstanding obligation from an earlier release.

## Evaluation Method

The evaluation follows the example through three stages. We first test release and repair histories in two storage representations. We then add a separately written implementation and generated variations to check whether the same obligations are realized consistently. Finally, we replace fixed scores with a trained recommender's outputs. A controlled comparison isolates the delivery consequence of treating a reporting omission as an authorization failure.

Each case specifies an execution and operations on its records, such as submitting permission evidence, deciding on delivery or proposing a correction. An operation is **accepted** if it can validly be recorded and **refused** if it violates a rule and leaves the history unchanged. Acceptance is therefore different from release: a valid publication record can state that the output is held. Likewise, an incomplete correction can be recorded as an unsuccessful attempt without closing the defect.

**Hand-authored histories.** Implementation C uses typed records and validates state transitions. Implementation L stores immutable log rows joined by reference constraints. The test set has 35 distinct histories under 38 coverage labels, covering valid lifecycles, single-defect variants, reporting-policy contrasts, and extra-input route and purpose checks. A separate audit reconstructs expectations and states from saved streams without importing either implementation. C and L nevertheless share authorization and basic data-processing helpers, so the next stage adds a separate realization of those rules.

**Generated cases and a third implementation.** Implementation I follows the same stated obligations without importing C, L or their helpers. Its authorization, record checks and refusal reasons are separately coded. The generated set contains 1,008 cases varying input route and purpose, whether inputs are reported, dependency depth, receipt/version agreement, permission state and reporting policy. Further variants add a second decision, a missing record reference or an audit event entered out of order. The realized factor levels, crossing pattern and changes from the initial protocol are documented in the reproducibility package.

For each case, the generator separately derives the expected authorization and delivery outcomes and submits corresponding records to each implementation. We compare those outcomes, reconstructed states and refusal reasons. We also check that omissions leave permission unchanged, invalid inputs prevent release, retries and refused operations append nothing, original records are preserved, closure follows the original assignment, and subsequent releases require renewed evidence. All implementations and the generator follow the same author-developed obligations. Agreement measures conformance to that model. The byte-preservation check additionally retains a shared harness helper, as documented in the package.

**Trained executor.** The first stages use a fixed output. We replace it with scores from a matrix-factorization recommender trained on goodbooks-10k, a public corpus of six million book ratings under CC BY-SA 4.0. Profile p0 supplies a personalized term, c0 a learned item bias, and x0 a term fitted on reading intent. These are controlled input roles whose responses generate both receipts and outputs. A dated amendment fixed settings, seed, input wiring and the comparison before data download. Training details and the substitution-script correction are retained in the package.

**Controlled comparison.** The alternative design treats any report/inventory discrepancy as an authorization failure and withholds the output under either reporting policy. It receives the same input evidence as our design. The comparison therefore identifies the delivery effect of combining these checks. It is also specified without a separate repair-authority assignment, although the measured delivery difference follows from its authorization rule.

# Results

## *Delivery Policy and Historical Responsibility*

Table 2 follows the two delivery policies through correction. In V2, release-with-repair delivers immediately and records the later correction. In V3, complete-report holds the result until correction, then requires a fresh decision basis before release. V5 extends the first history through a policy change, and V6 tests an incomplete correction. The histories show how delivery timing changes while responsibility for the report remains identifiable.

| Case | Reporting policy | Delivery and correction | Responsibility retained |
|---|---|---|---|
| V2 | Release with repair | Delivered first, correction verified later. | Original correction and closure assignment. |
| V3 | Complete report | Held until correction, then released using fresh evidence. | Correction does not replace the later permission check. |
| V5 | Release with repair, then complete report | Delivered once, correction remained pending through the policy change. | Original closure authority verified the correction. |
| V6 | Release with repair | Proposed correction was incomplete, leaving repair pending. | Failed verification did not close the defect. |

**Table 2. The principal release and repair histories observed in implementations C and L.**

The generated cases isolate the corresponding delivery contrast. Our design releases 56 outputs. The comparison withholds 28 of these because their reports omit permitted inputs and their policy allows release with repair pending. Another 28 cases are held by our design under complete-report. These two groups answer different questions: one identifies delivery lost by combining permission and completeness, while the other shows an intentional reporting-policy hold. Figure 1 places both groups within the generated set and shows which input and policy factors change the decisions. The counts describe the selected cases, not the frequency of omissions in deployed services.

## *Implementation Agreement and Evidence Checks*

C and L matched the expected decisions across all 38 coverage labels, and I reproduced those decisions, states and refusal reasons. In the generated set, all three implementations agreed on authorization and delivery in 1,008 cases and accepted the generator's expected decision records. Each processed 8,712 operation attempts, including 536 invalid attempts that appended nothing. Table 3 separates this decision agreement from the reason-code discrepancy and the follow-up probe.

The contrasts check specific parts of the design. Authorization was unchanged in 216 pairs differing only in report inclusion. All 672 cases with an input outside the allowed route or purpose were held. Of these, 168 isolate that condition, while 504 were already blocked by version mismatch or revoked permission. All 104 attempts to close a defect through a wrongly named authority were refused. Persistence of the original authority after a policy change is exercised separately by V5.

Receipt/version agreement provides another direct control. In 504 generated cases, the profile receipt identifies version 2 while the modeled resource and permission binding identify version 1. Authorization remains unresolved and delivery is held in all 504. Separate binding checks reject records referring to the wrong execution or input bundle. These observations concern consistency of the supplied evidence. The technical extension described below will test its connection to observed consumption.

The subsequent-decision cases contain 144 fresh and 144 stale decision bases. All 288 were accepted or refused as specified, but the frozen protocol predicted the wrong refusal reason for the 144 stale bases. It expected the recording-order check to fire, whereas the timestamp check fired first in every implementation. A separately amended probe preserved valid timestamps and changed the recording order in 144 fresh-basis cases. All three implementations then refused all 144 for the recording-order reason.

Thus decision agreement and reason-code agreement are reported separately, and the later probe directly exercises the check masked in the original cases.

## *Substitution of a Trained Recommender*

The trained recommender produced 1,008 distinct outputs, with 787 rankings different from the fixed-output version. Nevertheless, all three implementations retained the same governance outcomes in all 1,008 cases. This supports the intended implementation behavior: permission and reporting decisions depend on the input and policy evidence rather than the numerical recommendation scores.

On held-out ratings, the recommender's root-mean-square prediction error was 1.045, compared with 0.948 for a global-mean predictor, where lower is better. Its role here is to supply varying computed outputs for the governance test. The governance result is the unchanged decisions despite those output changes.

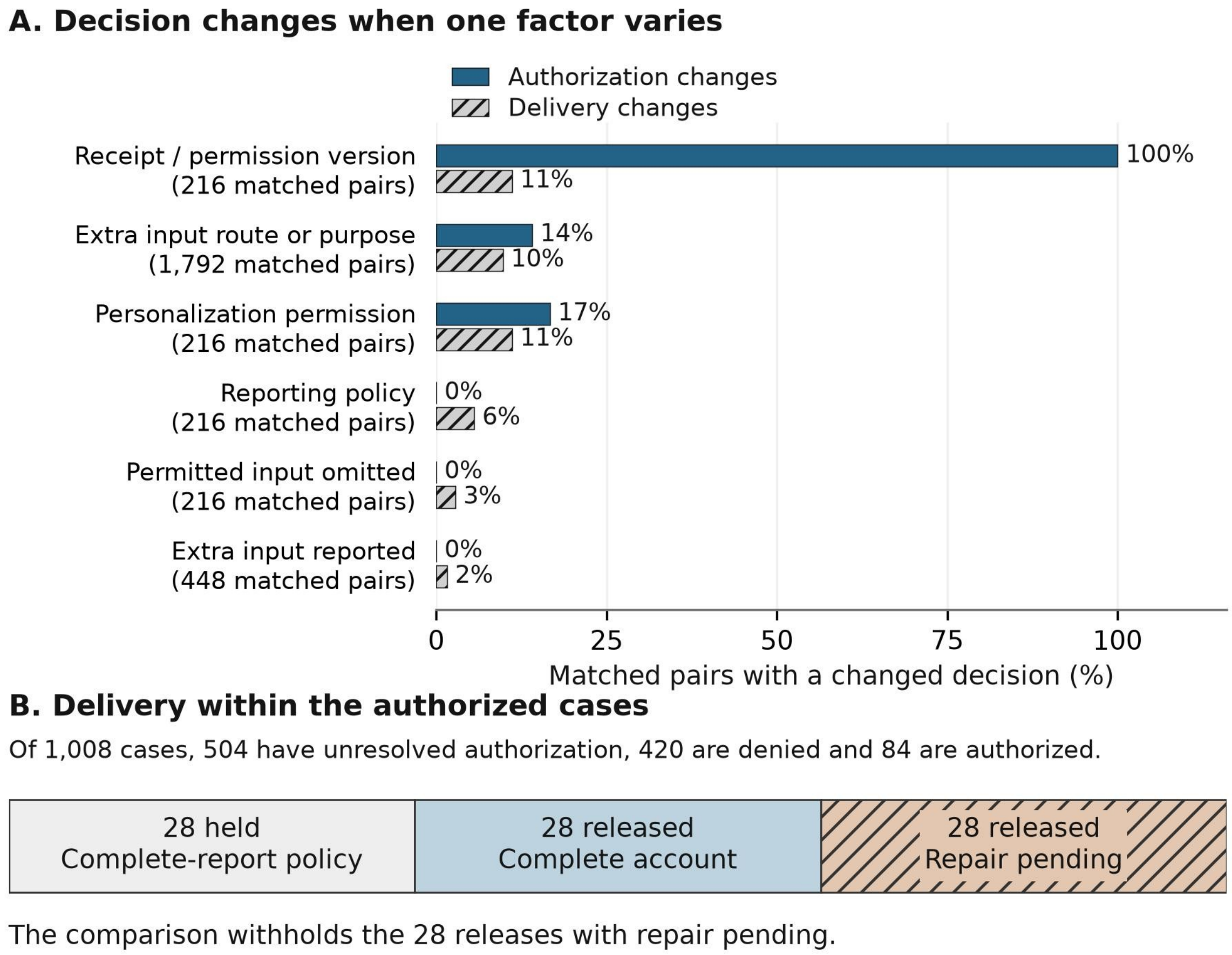

**Figure 1. Effects of input and policy variations in the 1,008 generated cases. Panel A compares pairs differing in one factor. Pair counts vary with the generator's crossing pattern. Panel B partitions authorized cases by reporting policy and report completeness.**

| Check | Cases or attempts | Observed result |
|---|---|---|
| C, L and I on generated decisions | 1,008 cases | Same authorization, delivery and reconstructed states. |
| Receipt version differs from authorization state | 504 cases | Authorization unresolved and delivery held. |
| Fresh versus stale subsequent-decision basis | 288 cases | All decision outcomes matched, but 144 refusal reasons differed from the protocol expectation. |

| Check | Cases or attempts | Observed result |
|---|---|---|
| Separate recording-order probe | 144 cases | All refused by the recording-order check in all implementations. |
| Trained scores replace fixed scores | 1,008 cases | Same governance outcomes, 787 rankings changed. |
| Invalid operation attempts | 536 per implementation | No records appended. |
| Comparison combining permission and completeness | 56 outputs our design releases | 28 withheld because their authorized inputs were incompletely reported. |

**Table 3. Decision agreement, evidence checks and the distinct reason-code result.**

## Discussion and Remaining Work

The artifact makes an organizational policy choice explicit. A service owner can require a complete account before delivery when documentation is a prerequisite, or allow delivery with correction pending when deferral is acceptable and correction and verification responsibilities are assigned. In a firm, the executor could be the model team, the publisher a release team, and the consent authority a privacy function. The interface connects their evidence to the owner's chosen delivery rule.

The same allocation guides disagreements about report completeness. The original report and received-input records remain available alongside the proposed correction. The assigned closure authority records whether that correction resolves the discrepancy. If policy allows closure without a corrected account, the original closure authority records that decision separately from verified correction. Such closure does not satisfy a complete-report requirement for delivery. Later delivery or a policy change does not transfer the original assignment. This makes explicit both the party expected to act and the party entitled to accept its account, providing a concrete basis for organizational accountability.

The evidence establishes implementation conformance in constructed histories with trusted identity, authority and receipt inputs, dependencies of one to three levels, and local persistence. A receipt identifies a received resource version. It does not by itself prove that the executor consumed that version or that the published result was computed from it. Binding results to observed consumption is therefore the next technical step, followed by testing whether practitioners can use the resulting evidence and responsibilities.

The planned extension has two stages. By mid-November, an instrumented executor would record consumed resource versions and bind that record to an execution identifier and output digest. Tests would substitute an input, alter an output, omit a receipt and supply an incorrect receipt to distinguish failures of execution evidence from report omissions. The target is an inspectable binding and fault cases that can be demonstrated at the workshop.

The second stage would invite 12-18 service owners and engineers with experience operating personalized services through professional networks. Participants would see contrasting V2, V3 and V5 scenarios, each presented once in counterbalanced order. They would be randomly assigned to linked decision-and-repair records or conventional chronological records containing the same facts. Participants would choose release or hold, identify who must correct the report and who may close the defect, then explain what a policy change alters. Two independent scorers would assess agreement with the stated policy and role assignments and code explanations for confusion between delivery and closure. Decision time and accuracy would be summarized by participant to retain the repeated-observation structure. Recruitment and piloting are targeted for November, with initial descriptive results and revised design requirements by early December. Together, the two stages would connect execution evidence to the decisions that service owners and engineers must make.

# Conclusion

Personalized services need to decide both whether an output may be delivered and who must correct an incomplete account of its inputs. Our interface separates input permission from the reporting policy governing delivery, then links the delivery record to the correction and verification responsibilities. In the evaluated histories, this permits delivery with repair pending when policy allows it. Those responsibilities persist even when reporting policy changes, while each later release requires renewed authorization evidence. The contribution is an executable allocation of decision rights that keeps current release decisions and earlier reporting obligations traceable across cooperating parties.

# Artifact Availability

The reproducibility package contains the three implementations, protocols and dated amendments, case generators, saved result streams, trained-executor configuration and audit scripts. A companion guide maps each reported result to its source and documents factor coverage and protocol changes. The package can be obtained from the authors.